\documentclass[12pt,hyper, notoc]{JHEP3}
\pdfoutput=1
\usepackage{amsmath,amssymb,calc}
\usepackage{bbm}
\usepackage[pdftex]{graphicx}
\usepackage{cite}
\usepackage{float}
\usepackage{epsfig}

\renewcommand{\thesection}{\arabic{section}}

\newcommand\be{\begin{equation}}
\newcommand\ee{\end{equation}}
\newcommand{\bea}{\begin{eqnarray}}
\newcommand{\bean}{\begin{eqnarray*}}
\newcommand{\eean}{\end{eqnarray*}}
\newcommand{\eea}{\end{eqnarray}}

\def\beq{\begin{equation}}
\def\eeq{\end{equation}}
\def\id{\protect{{1 \kern-.28em {\rm l}}}}

\def\unit{\relax{\rm 1\kern-.26em I}}

\title{Maxwell-Chern-Simons Vortices and Holographic Superconductors}

\author{ Gianni Tallarita and Steven Thomas \\ \\ 
Queen Mary University of London \\
Center for Research in String Theory \\
Department of Physics, \\
Mile End Road, London, E1 4NS, UK. \\ 
\\
Email: \email{G.Tallarita@qmul.ac.uk, S.Thomas@qmul.ac.uk}}

\abstract{We investigate vortex solutions of a charged scalar field in Einstein-Maxwell theory in 3+1 dimensions with the addition of an axionic coupling to the Maxwell field. We show that the inclusion of such a term, together with a suitable potential for the axion field, can induce an effective Chern-Simons term on the 2+1 dimensional boundary.  We obtain numerical solutions of the equations of motion and find Maxwell-Chern-Simons like magnetic vortex configurations, where the magnetic field profile varies with the size of the effective Chern-Simons coupling. The axion field has a non-trivial profile inside the AdS bulk and on the 2+1 dimensional boundary but does not condense at spatial infinity.  }

\preprint{}
\begin{document}

\section{Introduction}
The subject of holographic superconductivity (HS) has received much attention in the past 2 years since it was initially proposed in \cite{Hartnoll}, \cite{Hartnoll2} (see \cite{Horowitz} for a recent review),   \cite{Roberts3} - \cite{horowitz4}
 It has provided a fascinating and novel approach to the study of phase transitions in certain 2+1 dimensional systems by embedding them holographically in the background of an asymptotically AdS black hole. Phase transitions in the boundary theory appear in  this setting as scalar hair that develops below a certain critical temperature when a charged scalar field is  added to Einstein-Maxwell theory which supports either an AdS Schwarzchild or Reissner-Nordstrom type black hole solution.  This bulk scalar field  is known to be dual to  certain operators in the boundary CFT and the development of scalar hair outside the black hole corresponds to these operators `condensing', signifying a phase transition.

Strictly speaking, in standard HS \cite{Hartnoll},\cite{Hartnoll2}, the dual CFT has no dynamical gauge field, so the phase transition is one that appears to describe the change in the superfluid  phase rather than a superconducting phase. \footnote{ A very interesting recent paper \cite{Pomarol2} discusses how the standard model of a HS can be modified to include a dynamical gauge field in the boundary CFT.} 

 In \cite{Johnson} a different approach was taken by allowing for a  $z $-independent term in the expansion of $A_\chi (r, z) = 
a_\chi (r) + z J_\chi(r)+ O(z^2) $.  Here $(r,\chi )$ correspond to 2d polar coordinates of the plane at fixed $z$, where $z$ is the conformal coordinate of the `intetrior' of the  $AdS_4$ background, with $z=0$ corresponding to the 2+1 dimensional boundary at infinite distance from the AdS black hole. The expression $\frac{1}{r} \partial_r a_\chi (r) $ is then identified with a boundary magnetic field which one can think of  as the result of  weakly gauging the global $U(1)$ symmetry of the boundary CFT.
The resulting solutions are rather non-trivial and depending on the various assumptions and boundary conditions imposed, were shown to describe either magnetic droplets  or magnetic vortices living on the boundary (the authors of \cite{Pomarol} obtained similar solutions although using a different approach where a magnetic field is added by hand on the boundary at $z=0$ and its flux constrained to be quantised and related to the quantised superfluid vorticity).

Holographic supefluid vortices have also been studied in \cite{keranen1 },\cite{keranen2  }. Here the approach  is to take $a_\chi(r)=0 $ so there is no magnetic field. Solutions to the bulk Einstein-Maxwell theory were found which on the  CFT side describe instead superfluid vortex configurations. In this solution the superfluid velocity is  related to the current $J_\chi$ appearing  in the expansion of the $A_\chi  $ component of the bulk $ U(1) $ gauge field near the CFT boundary at $z=0$.   
 
The appearance of such magnetic vortices (or superfluid vortices) on the 2+1 dimensional boundary at asymptotic infinity are triggered when the effective temperature of the system is above  the critical value  for which the scalar field condenses. They have quantised magnetic flux (or vorticity) and are thus topologically stable and are very reminiscent of the vortices discovered by Nielsen and Olesen in flat 2+1 dimensional space-time \cite{Nielsen}. 
 
There  exist, however, different kinds of magnetic vortices in flat 2+1 dimensional space-time if one includes not just a Maxwell term but also a  Chern-Simons (CS)  term for the electromagnetic potential. Pure  self-dual Chern-Simons vortices exist (when no Maxwell term is present) for suitably chosen charged scalar field potential \cite{HKP}, \cite{Jackiw}. These pure CS vortex solutions are topologically stable, have quantised magnetic flux but in addition, unlike pure Maxwell type vortices, they also carry quantised electric flux.  A distinguishing characteristic feature is a magnetic field that peaks in a ring  outside of the `core', unlike the pure Maxwell case where the magnetic field peaks inside the core. It is also possible to consider topological magnetic vortices in a  theory which has both Maxwell and a Chern-Simons term present for the $U(1) $ gauge field (see eg \cite{Zhang} for a comprehensive recent review). It is an interesting problem to see if such  CS or mixed Maxwell-CS  vortices can appear in a holographic setting.

In the `standard' model describing HS  a complex scalar $\Phi$ is coupled to a $U(1)$ gauge field described by a standard Maxwell action on an $AdS_4$ background. This scalar has a negative mass $V(\Phi)=-\frac{2}{L^4} |\Phi|^2$, where $L$ is a scale defined by a non-vanishing cosmological constant $\Lambda$, which remains above the Breitenlohner-Freedman stability bound \cite{Breiten} and thus does not induce an instability of the system. 

 The obvious  problem one faces in trying to generalise the dynamics of the bulk gauge field  to include a CS action is that no CS term exists in a 3+1 dimensional theory. 
\footnote{gravity duals in 4+1 dimensions with both Maxwell and Chern-Simons terms present have been considered in \cite{nakamura}, \cite{Ooguri}} 
In principle one could add by hand a boundary action containing a pure CS term for the gauge field. However such an approach will lead to the appearance of explicit delta functions in the bulk equations of motion. Whilst the solution of this new system of equations may still be technically feasible, it is in some sense against the spirit of holography and we will not pursue this approach here.
     
 Instead we shall adopt  a more indirect approach.  As we will see,  by including an additional pseudo-scalar axion field, $\theta$, in our $AdS_4$ background, with  the usual coupling to the topological term  $F\wedge F $, we can induce an effective CS term for the  $U(1)$ gauge fields that live on the  2+1 dimensional boundary.  Such a term would then modify the magnetic vortex solutions found in \cite{Johnson}, \cite{Pomarol} and  perhaps display properties of the more general  flat space Maxwell-CS vortices mentioned above such as the magnetic field peaking outside of the core. The coupling $\theta F\wedge F $ can arise, for example, due to the axial anomaly if one included charge chiral fermions in the bulk theory. 

This paper is organised as follows: in section 2 we introduce the Maxwell theory with charged scalar in the $AdS_4$ background, including, in addition,  the axion field and its couplings.  We make an ansatz for the dynamical fields, determine their equations of motion and discuss the relevance of terms in their near-boundary $z=0$ expansion. We shall see that the inclusion of the $\theta F \wedge F  $ term gives rise to an effective CS term on the boundary.
Section 3 describes numerical solutions of this system, where we show that below a critical temperature the scalar field condenses and the corresponding magnetic field profile varies with the parameter $\kappa$, (the coefficient in front of the $\theta F\wedge F $ term, which is related to the effective CS coupling on the boundary). 
In this section we also discuss the azimuthal current $J_{AdS}$, the charge density $\rho_{AdS}$ and the axion field $\theta (r)$ evaluated on the boundary and investigate the $T>T_c$ region, where we show that the system admits a ``trivial" solution where the scalar field vanishes asymptotically. In Section 4 we perform a near-boundary expansion of the equations of motion to try and uncover the asymptotic analytic properties of the numerical solutions for large and small distances in the 2+1 dimensional boundary . Finally, in Section 5, we provide a short discussion of the results and directions for further research. 
 
\section{The Model} 

The action we consider is:
\be
S=\int d^4x\sqrt{-G}\left(\frac{1}{16\pi G_N}(R-2 \Lambda)-\frac{1}{g^2}\mathcal{L}\right),
\ee
where
\be\label{action1}
\Lambda = - \frac{3}{L^2}, \mathcal{L}=\frac{1}{4}F_{\mu\nu}F^{\mu\nu}+\frac{1}{L^2}|D_\mu \Phi|^2-\frac{2}{L^4}|\Phi|^2+\frac{\kappa}{\sqrt{-G}}\theta \epsilon^{\mu\nu\rho\sigma}F_{\mu\nu}F_{\rho\sigma}+(\partial_\mu \theta)^2+V(\theta)
\ee

where $G_N$ is the gravitational coupling, $g$ the gauge coupling, $D_\mu=\partial_\mu-iA_\mu$ and $V(\theta)$ is the axion potential. The constant $\kappa$ is, as we shall soon see, related to an effective  CS coupling  on the boundary. In our coordinate notation,  $z$ denotes a ``radial'' or ``bulk" direction of the $AdS_4$ space with $z=0$ being its boundary at infinity and $(r,\chi)$ acting as coordinates for two-dimensional planes of constant $z$. The above form of the action in which the gauge coupling $1/g^2$ appears only as an overall multiplicative factor in front of  the `matter' Lagrangian is useful in working in the probe limit (which we shall assume) where $g$ is taken very large. To bring the action into this form, the canonical fields in the Lagrangian are redefined as 

\be
A_\mu\rightarrow\frac{1}{g}A_\mu,\quad \Phi\rightarrow \frac{1}{g} \Phi, \quad
\theta\rightarrow\frac{1}{g}\theta,\quad \kappa\rightarrow g\kappa
\ee

These rescalings allow us to work in the formal decoupling limit of $g\rightarrow\infty$ where one can safely assume that the background space is fixed, uncharged and suffers from no back-reaction from fields in the Lagrangian. The space time is the $AdS_4$-Schwarzschild black hole described by the metric

\bea
ds^2&=&\frac{L^2}{z^2}(-f(z)dt^2+dr^2+r^2d\chi^2)+\frac{L^2}{z^2f(z)}dz^2\nonumber \\
f(z) &= &1-z^3 \nonumber\\
\sqrt{-G}&=&\frac{L^4r}{z^4}
\eea

with $z=1$ the location of the black hole horizon. This has a Gibbons-Hawking \cite{Gibbons} temperature $T$, which is dual to the $CFT$ temperature $T$ via the $AdS/CFT$ dictionary. For the purpose of a HS one is interested in the dependence of a dimensionless measure of the condensate as a function of a dimensionless measure of the temperature (as explaine din \cite{Horowitz} ), which we shall discuss later in Section 3. \newline


We work with an ansatz of the form
\bea
A_z&=&0\nonumber\\
A_r&=&0\nonumber\\
\Phi&=&\phi(r,z)e^{in\chi}\nonumber\\
A_\chi&=&A_\chi(r,z)\nonumber\\
A_0&=&A_0(r,z)\nonumber\\
\theta&=&\theta(r,z)\nonumber\\
V(\theta)&=&m^2_\theta\theta^2+\lambda\theta, \qquad V(\Phi) = -\frac{2}{L^4} |\Phi|^2
\eea

Note in the above we have included the possibility of a  linear term on the axion potential.  The presence of the axionic coupling  $\theta F\wedge F $ already violates $\theta\rightarrow -\theta$ parity so in principle such a term could be included. However as we shall see later, the coefficient $\lambda$ of the linear term must be related to the value of $\theta $ evaluated on the boundary at $z=0$
in order to solve the equations of motion. This necessarily means that taking  $\lambda  \neq 0 $,  $\theta$ must be  a constant everywhere on the boundary and so has a trivial profile there. It is important to point out that this constant value of $\theta$  should not be thought of as  the axion field condensing in this system. This is because any constant appearing in $\theta$ can be effectively removed by adding an explicit term $F\wedge F $ to the bulk action with an appropriate coefficient. Thus no axionic ``hair"  exists. This is consistent with general arguments  
(see e.g. \cite{Horowitz}) that the $AdS_4$ black hole can only develop hair from charged scalar fields.
\footnote{ Note that black hole solutions  in 3+1 dimensions are known to exist which do indeed develop `conventional' axionic hair (by conventional we are excluding the possibility of stringy axionic hair that can arise when one has a  dilaton and Kalb-Ramond field present  which arise from a string theory compactification) when one includes the term $\theta F \wedge F $ \cite{Reuter}, \cite{Duncan}. However such hair only arises when one includes torsion into the background geometry. }\newline

With the above ansatz we obtain the following bulk equations of motion
\begin{align}\label{equations1}
&\frac{-\sqrt{-G}z^4}{L^4f}\left(f\partial_z^2A_0+\frac{1}{r}\partial_r(r\partial_rA_0)-\frac{2A_0}{z^2}\phi^2\right)-8\kappa\epsilon^{0\nu\rho\sigma}(\partial_\nu\theta)F_{\rho\sigma}=0\nonumber
\end{align}
\begin{align}
&\frac{\sqrt{-G}z^4}{L^4r^2}\left(\partial_z(f\partial_zA_\chi)+r\partial_r\left(\frac{1}{r}\partial_rA_\chi\right)-\frac{2}{z^2}\phi^2(A_\chi-n)\right)-8\kappa\epsilon^{\mu\nu\rho\chi}F_{\nu\rho}\partial_\mu\theta=0\nonumber
\end{align}
\begin{align}
&z^2\partial_z\left(\frac{f}{z^2}\partial_z\phi\right)+\frac{1}{r}\partial_r(r\partial_r\phi)+\left(\frac{A_0^2}{f}-\frac{(A_\chi-n)^2}{r^2}-\frac{m^2}{z^2}\right)\phi=0\nonumber
\end{align}
\begin{align}
&z^2\partial_z\left(\frac{f}{z^2}\partial_z\theta\right)+\frac{1}{r}\partial_r(r\partial_r\theta)-z^2\kappa F\tilde{F}+\frac{1}{z^2}\frac{dV}{d\theta}=0,
\end{align}

In what follows we will be interested in the behaviour of the various fields near the boundary $z=0$. We shall assume the fields have the following expansions for small  $z$

\bea\label{expansions}
\phi(r,z)&=&z\phi_0(r)+z^2\phi_1(r)\nonumber\\
A_0(r,z)&=&\mu+z\rho+z^2A_2+ z^3A_3 +z^4 A_4.+..\nonumber\\
A_\chi(r,z)&=&a_\chi +zJ_\chi+z^2D_2+....\nonumber\\
\theta(r,z)&=&\theta_0(r)+z\theta_1(r)+z^2\theta_2(r)+...
\eea

The $AdS/CFT$ correspondence \cite{Maldacena} teaches us that terms in the near boundary expansion of bulk fields act as sources for conformal operators in the dual theory. More precisely, $\phi_i$ sets the vacuum expectation value of an operator $O_i$ with dimension $i$. For consistency, only one of these operators can exist at one time so focussing on the $\phi_1$ case will exclude $\phi_2$ and vice-verse. We will only consider $\phi_1$ in the following. Within the context of 
HS the scalar field describes, in the dual CFT, the order parameter for the spontaneous breaking of the $U(1)$ symmetry. Similarly the first and second terms in the expansion of $A_0$ are associated to the Maxwell contribution to the chemical potential and charge density respectively, whilst those of $A_{\chi}$ are related to the magnetic field $B=\frac{1}{r}\partial_r a_\chi $ and azimuthal current density $J_\chi$. With the presence of a Chern-Simons boundary term the precise definition of these quantities acquires a further contribution. For the detailed relations we refer the reader to the Appendix.

\section{Solution}

We are interested in solving the coupled partial differential equations eq(\ref{equations1}). In order to avoid divergences at $r=0$ we make the following field redefinitions 
\bea
\phi&=&zr^n\tilde{\phi}\nonumber\\
A_\chi&=&r^2\tilde{A_\chi},
\eea

such that, near the boundary,

\bea\label{Atheta}
\tilde{\phi}&=&\tilde{\phi}_0+z\tilde{\phi}_1\\
\tilde{A}_\chi&=&\tilde{A}_0+z\tilde{A}_1+z^2\tilde{A}_2+...
\eea

This implies that the magnetic field becomes
\be\label{magfield}
B = 2\tilde{A}_0+r\partial_r\tilde{A}_0.
\ee

and the condensate is
\be
\phi_1 (r) = \frac{1}{2} \partial_z^2\phi = r^n\tilde{\phi}_1.
\ee

With these redefinitions and setting $L=1$, the equations of motion become
\begin{align}\label{equations}
\partial^2_z\tilde{\phi}+\frac{f'}{f}\partial_z\tilde{\phi}+&\frac{1}{f}\left(\partial^2_r\tilde{\phi}+\frac{2n+1}{r}\partial_r\tilde{\phi}+\frac{n^2}{r^2}\tilde{\phi}-\frac{1}{r^2}\tilde{\phi}(r^2\tilde{A_\chi}-n)^2\right)\nonumber\\
&+\frac{1}{f^2}\tilde{\phi}A_0^2+\left(\frac{f'}{zf}-\frac{2}{z^2}+\frac{2}{z^2f}\right)\tilde{\phi}=0
\end{align}
\begin{align}
\partial^2_z\tilde{A_\chi}+\frac{f'}{f}\partial_z\tilde{A_\chi}+&\frac{1}{f}\left(\partial^2_r\tilde{A_\chi}+\frac{3}{r}\partial_r\tilde{A_\chi}\right)-\frac{r^{2n}}{f}\tilde{\phi}^2\left(\tilde{A_\chi}-\frac{n}{r^2}\right)\nonumber\\
&-\frac{16\kappa}{rf}(-\partial_r\theta\partial_zA_0+\partial_z\theta\partial_rA_0)=0
\end{align}
\begin{align}
\partial^2_zA_0+\frac{1}{f}\left(\partial^2_rA_0+\frac{1}{r}\partial_rA_0\right)-\frac{r^{2n}}{f}\tilde{\phi}^2A_0\nonumber\\
+\frac{16\kappa}{r}(-r^2\partial_r\theta\partial_z\tilde{A_\chi}+r^2\partial_z\theta\partial_r\tilde{A_\chi}+2r\partial_z\theta\tilde{A_\chi})=0
\end{align}
\begin{align}
z^2\partial_z\left(\frac{f}{z^2}\partial_z\theta\right)+\frac{1}{r}\partial_r(r\partial_r\theta)+\frac{1}{z^2}(2m_\theta^2\theta+\lambda)&\nonumber\\
+\frac{8\kappa z^2}{f}(-r^2\partial_rA_0\partial_z\tilde{A_\chi}+r^2\partial_r\tilde{A_\chi}\partial_zA_0+2r\tilde{A_\chi}\partial_zA_0)=0
\end{align}

We are interested in solving these equations to extract the boundary behaviour of the fields. Unsurprisingly the equations cannot be solved analytically, they do however present numerical solutions. 

Before investigating these, an important observation to make  is that by comparing the  $A_0$  or $\tilde{A}_\chi$ equations of motion with those derived from a purely 2+1 dimensional Maxwell-CS theory with charged scalar matter,  we see the term $\kappa \partial_z \theta  \sim  \kappa \theta_1(r)$ (whose origin is the axion coupling $\theta F\wedge F$ ) generates an effective  CS coupling  in the 2+1 dimensional boundary theory at $z=0$.
\subsection{Solution 1}

For the numerical procedure, we must specify boundary conditions on the fields. Initially, we work with $\mu=const$, $J_\chi=const$, $\theta_0=1$ and $n=1$. Then, for consistency at the boundary $z=0$, one finds $m^2_\theta=1$ and $\lambda=-2$. In all solutions found, the scalar field (Figure 1) condenses independently of the parameter $\kappa$. This is in close agreement with the results of \cite{Pomarol} in which the scalar field vanishes at the location of the vortex. In Figure 2 we present the results for the magnetic field $B(r)$. We find that for low enough $\kappa$ the system is dominated by a pure Maxwell solution, where the magnetic field peaks at the origin and vanishes for large $r$ \cite{Johnson}. However, as $\kappa$ is increased the solution is driven away from the pure Maxwell configuration until eventually, at a critical value of $\kappa$ the magnetic field vanishes at the origin and peaks at finite distance from it (red line), once again decaying to zero at large $r$. This behaviour is characteristic of self-dual Chern-Simons vortices in flat space \cite{Jackiw} where $B(r)$ forms a ring surrounding the origin. At even higher values of $\kappa$ the magnetic field becomes negative close to the origin. The reader might at this point be worried about flux quantization, however the flux $\Phi_B =\int r B drd\chi$ always remains positive due to uneven contributions to the integral from below and above the zero axis, hence we expect to maintain flux quantisation as required for topological solitonic solutions. The Charge density, which depends on the  parameter $\kappa $ is plotted in  Figure 3. It  tends to a constant value at asymptotically large values of $r$  approximately independent of the value of $\kappa$, at least for the values chosen.  The effect closer to the core of the vortex, which appears to be a depletion of electric charge is in fact a temperature dependent phenomenon. This was first found in \cite{Johnson} where the authors showed that upon lowering the temperature in the region $T<T_c$ the charge density switches from increasing to decreasing there. This effect, if not simply due to the complex numerical procedure, was suggested to be indicative of strong screening effects from the core of the vortex. However one should not interpret this as signifying that the vortex cannot support electric charge, as indeed one expects a Chern-Simons vortex to do.\\

\begin{figure}[p]
\label{phifield}
\includegraphics[width=1.5\textwidth,height=0.4\textheight]{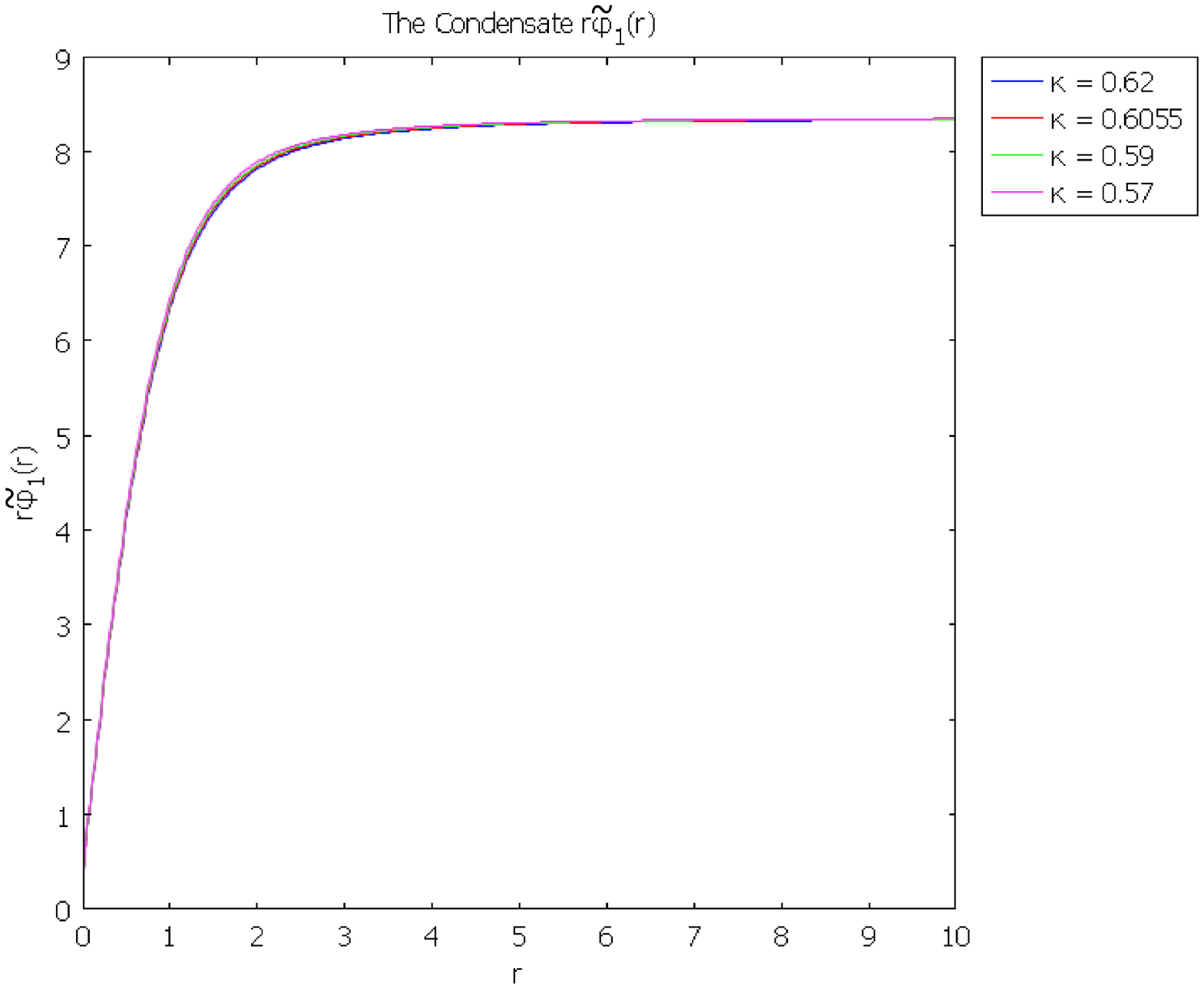}
\caption{The Condensate $\phi_1(r) = r\tilde{\phi}_1(r)$ (in units of $\mu = 1$) for different values of the  parameter $\kappa$. As expected, the condensate vanishes where the vortex is placed. There is negligible variation in the condensate profile for this range of $\kappa$.}
\end{figure}
\begin{figure}[p]
\label{Bfield}
\includegraphics[width=1\textwidth,height=0.35\textheight]{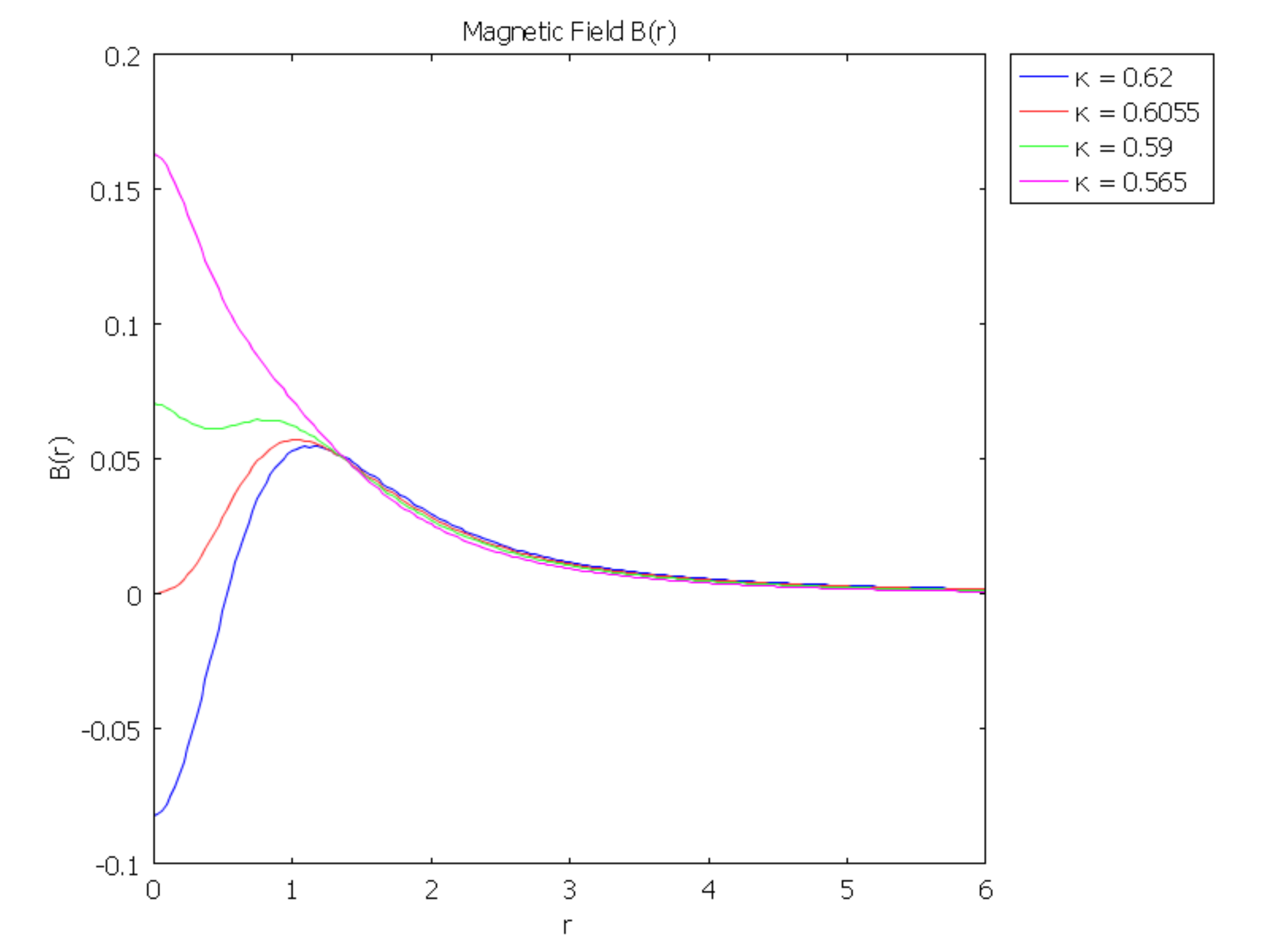}
\caption{The Magnetic Field $B (r)$ for different values of the parameter $\kappa$. For low values of $\kappa$ with the magnetic field peaks at the origin (purple line) as per the pure Maxwell case. As one increases $\kappa$ this profile is modified (green line) until at a critical $\kappa = 0.6055$ we see that the Magnetic Field vanishes at the origin and forms a ring around it (red line). For increasing $\kappa$ the magnetic field falls below the zero axis.}
\end{figure}
\begin{figure}[p]
\label{rhofield}
\includegraphics[width=1\textwidth,height=0.35\textheight]{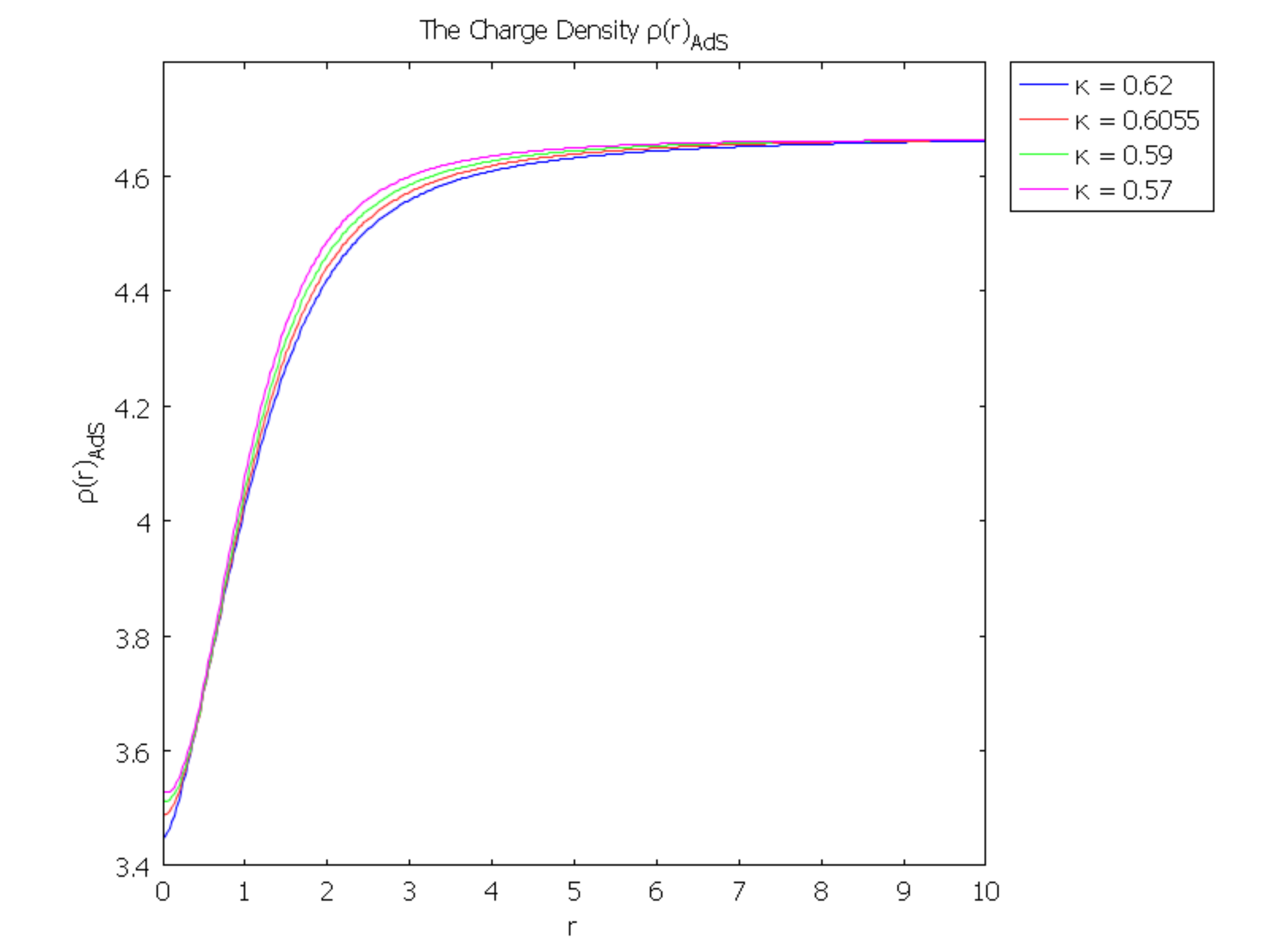}
\caption{The Charge Density $\rho_{AdS}(r)$ for different values of the  parameter $\kappa$. The charge density tends to a constant for large $r$ and shows little modification with varying values of $\kappa$.}
\end{figure}
It is also interesting to study the behaviour of the axion field $\theta_1(r)$. In Figure 4 we show plots for different values of $\kappa$. Increasing the parameter $\kappa $ pushes the axionic field to lower values whilst preserving its shape. 
The importance of  $\theta_1(r)$ is that for large $r$ it plays the role of a CS coupling on the  boundary, as discussed earlier. As we see from Figure 4, $\theta_1 $  tends to a constant  for large $r$, which lends further weight to this interpretation.
\begin{figure}[p]
\label{thetafield}
\includegraphics[width=1\textwidth,height=0.35\textheight]{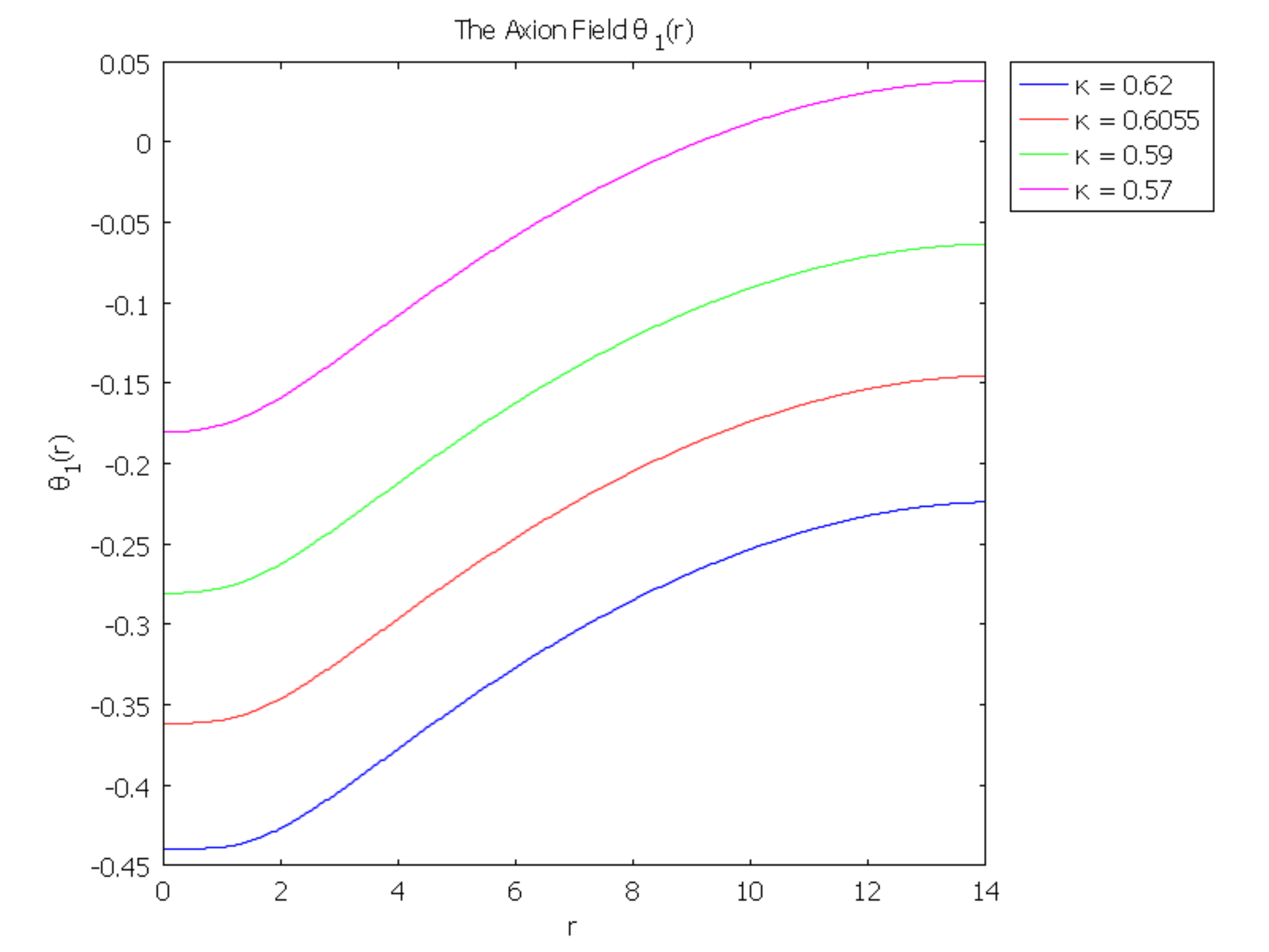}
\caption{The Axion Field $\theta_1(r)$ for different values of  $\kappa$. The Axion Field is negative and increases with smaller values of $\kappa$.}
\end{figure}

In passing we also mention there exists another set of solutions for which, rather than imposing Dirichlet conditions on $\mu$, we allow for this to be dynamically determined by the numerical procedure whilst keeping $\rho=const$. Then by having a non-constant $\mu$ at the boundary one can observe a pure Chern-Simons contribution to the azimuthal current $J_\chi$, which was previously determined only by it's pure Maxwell component set manually to a constant (see Appendix).  
We leave the numerical  exploration of such solutions, and the corresponding profile for the modified azimuthal current to future work \cite{TT}.
\subsection{Solution 2}
\begin{figure}[p]
\label{condformu}
\includegraphics[width=1\textwidth,height=0.35\textheight]{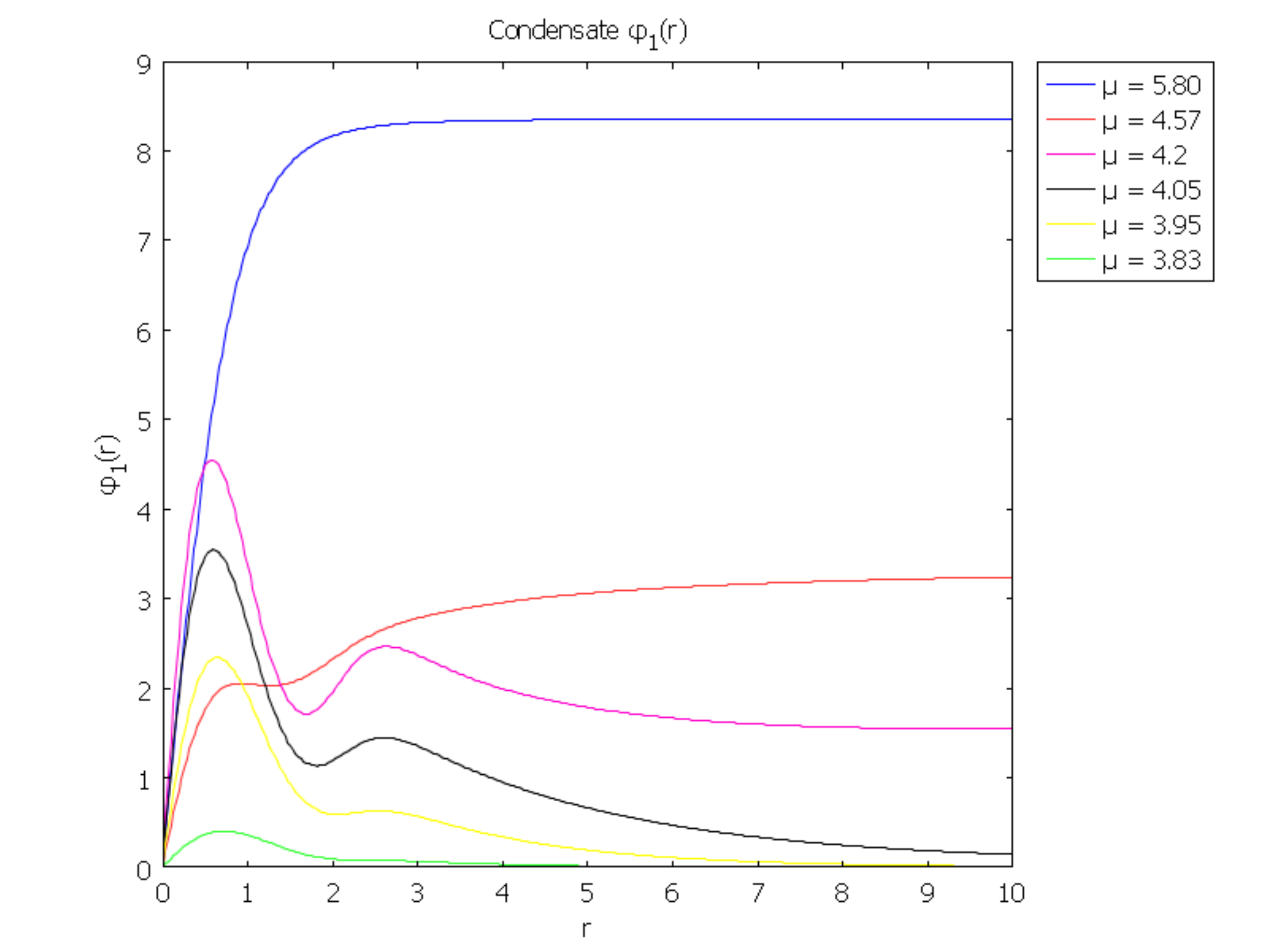}
\caption{The condensate $\phi_1(r)$ (in units of $\mu = 1$) for different values of the chemical potential $\mu$, and hence different temperatures. One sees that below $T_c$ the scalar field condenses, but as one raises the temperature the scalar field drops gradually to zero.}
\end{figure}
This section is devoted to analysing the set of solutions for $T>T_c$, the critical temperature at which the scalar field condenses. As explained in \cite{Horowitz} we are interested in the dependence of the condensate as a function of a dimensionless measure of the temperature. We take $\frac{T}{\mu}$ for this measure, where $T$ is simply the Gibbons-Hawking Temperature of the black hole and remains constant for varying parameters of the numerical procedure. Hence, varying values of $\mu$ allow us to change the effective temperature of the $CFT$. Figure 5 shows the profile of the condensate for different values of $\mu$. As we make $\mu$ smaller, and hence the effective Temperature larger, the solution enters the region of $T>T_c$ and the scalar field doesn't condense. For increasingly small $\mu$, the solution is driven towards $\phi_1 =0$. Conversely, when one raises $\mu$, the profile of the solution remains that of the condensing scalar (blue line) as the system is well within the $T<T_c$ region.

\section{Asymptotic analysis}

Even though the set of coupled equations eq(\ref{equations}) is highly non-linear and can only be solved numerically, one can try and  perform an asymptotic analysis to get an analytical feel for the behaviour of the functions at the $z=0$ boundary. In the case of purely 2+1 dimensional CS vortices or mixed Maxwell-CS vortices, such an analysis proves useful in  providing analytic expressions for the various fields in the regions near the core and far away from the vortex (see e.g.\cite{Zhang} for a detailed analysis). These analytic expressions, apart from their own intrinsic interest, can provide further checks on any full numeric solution.

In our model we  first need to obtain a series of equations  which are  derived from the full equations of motion by taking a small $z$ expansion. For the origin of the various fields appearing in this expansion, we refer the reader to eq(\ref{expansions}). \
From the equation of motion of the scalar field, expanding the fields around $z=0$ using eq(\ref{Atheta}) and eq(\ref{expansions}), at $O(z)$ we find
\be\label{A0}
2n\tilde{A}_0\tilde{\phi}_1-r^2\tilde{A}_0^2\tilde{\phi}_1+(1+2n)\frac{\partial_r\tilde{\phi}_1}{r}=0.
\ee

\noindent Similarly, the $A_0$ equation gives
\be
\partial_r\rho\left(\frac{1}{r}+16r\kappa\theta_1\right)+\partial_r^2\rho+16\kappa r\rho\theta_2+A_3=0,
\ee

\noindent at order $O(z)$, and 
\bea\label{asymphi}
-2r^{(1+2n)}\mu\tilde{\phi}_1^2+\partial_r(r\partial_rD_2(r))+16\kappa \theta_1\left(\partial_r(r^2D_2(r))\right)-32\kappa \partial_r\theta_1r^2 D_2(r)\nonumber\\
+16\kappa r\tilde{A}_1(4\theta_2+r\partial_r\theta_2)+rA_4+16\kappa r\theta_3(2\tilde{A}_0+r\tilde{A}_1)=0
\eea

\noindent at order $O(z^2)$. These equations govern the behaviour of the magnetic field, the condensate and the charge density shown in the solutions above. They are a complex system of coupled partial differential equations. In particular, with a simple rearrangement eq(\ref{A0}) can be expressed as
\be
\tilde{A_0}=\frac{n}{r^2}+\frac{1}{r^{3/2}}\sqrt{\frac{n^2}{r}+(1+2n)\frac{\partial_r\tilde{\phi}_1}{\tilde{\phi}_1}},
\ee

\noindent which implies that the magnetic field eq(\ref{magfield})

\be\label{asymb}
B=\frac{(1+2n)(-r(\partial_r\tilde{\phi}_1)^2+\tilde{\phi}_1(\partial_r(r\partial_r\tilde{\phi}_1)))}{2r^{3/2}\tilde{\phi}_1^2\sqrt{\frac{n^2}{r}+(1+2n)\frac{\partial_r\tilde{\phi}_1}{\tilde{\phi}_1}}}.
\ee

Then, to observe the full dependence of the magnetic field on $\kappa$, one would have to substitute eq(\ref{asymphi}) rearranged for $\tilde{\phi}_1$ in eq(\ref{asymb}), even for $\kappa \neq 0$ this is easy to do but, for the sake of simplicity, we will not quote the full result. Instead, for the simplifying choice of $\kappa=0$, where one has pure Maxwell, note that one can read off the asymptotic form of the condensate, and eq(\ref{asymphi}) becomes

\be\label{phi1}
\phi_1(r) = r^n\tilde{\phi}_1=\sqrt{\frac{\partial_r(r\partial_rD_2(r))+rA_4(r)}{2\mu r}},
\ee

so the condensate contains functions appearing in the expansion of $A_\chi$ and $A_0$ (recall that in our numerical simulation $\mu$ is kept constant). 

Indeed the lesson we are learning is that the set of equations that arise at each order in $z$ do not form a closed set, in that  the number of functions is always  greater than the number of equations. Even at lowest order we see from the above that additional 'source'  terms $D_2(r), A_4 (r) $  appear in the 3 equations that determine $\tilde{A}(r), B(r) $ and $\tilde{\phi}_1 (r)$ and from which one could try and extract the small and large $r$ behaviour. But to do so requires some knowledge of the behaviour of the functions $D_2(r), A_4(r)$ in these same limits. 

Thus the situation is, unfortunately rather more complex than in the case of strictly 2+1 dimensional  Maxwell-CS vortices \cite{Zhang}. One way of proceeding is to make approximate analytic fits to the functions $D_2(r) $ and $ A_4(r)$ from
their numerical profiles (which can be derived though we have not presented the plots here) for small and large $r$. Substituting these into eq(\ref{phi1}) would then allow one to extract analytic asymptotic expressions for the condensate
$\tilde{\phi}_1 $,  $B(r) $ and $\tilde{A} (r)$. In doing this one would impose the appropriate boundary conditions  on the latter fields for $r \rightarrow 0$ and $r \rightarrow \infty$  in order to to yield finite energy solutions on the 2+1 dimensional boundary at $z=0$. We will leave this analysis  to future work
\cite{TT}.

\section{Discussion}

In this paper we have made  a first attempt to find a holographic description of magnetic vortices in 2+1 dimensions where the the abelian gauge field is  governed by both a Maxwell term and an effective CS term. We have achieved this within a 3+1 dimensional theory  describing an AdS black hole
(in the probe limit), by including an additional neutral scalar field, the axion, and its coupling to the topological term $F\wedge F$. 
The resulting equations of motion are, like the original HS, rather difficult to solve analytically and we  found instead numerical solutions for the magnetic field, scalar condensate, charge density and axion fields as functions of the coordinates $(r,z)$.  The corresponding profiles of these fields on the boundary at $z=0$ were presented. For a range of values of the chemical potential $\mu $ corresponding to $T<T_c$ they clearly describe  magnetic vortex like 
solutions. An important difference when compared to those found in the literature \cite{Johnson}, \cite{Pomarol}, is that the magnetic vortices we have found have the magnetic field peaking outside of the core when the parameter $\kappa$ that multiplies the $\theta F\wedge F$ term in the action,
 (and which is  related to the effective CS coupling on the boundary), is above a certain value.
 Such a property of the magnetic field `peaking' in a ring outside the core is very reminiscent of the profile of a magnetic field outside of a CS vortex \cite{HKP}, \cite{Jackiw}.

By lowering the values of $\mu$ we have also  explored the region $T >T_c$ and have found numerical profiles that show the  condensate gradually vanishing at large distances as one decreases $\mu$.
 
A further novel behaviour we have observed is the field reversal of the magnetic field near to the vortex core  (in the case $T<T_c$) as  one continues to increase the parameter $\kappa $. Whilst the phenomenon of distance dependent field reversal in magnetic vortices is known to occur in anisotropic superconductors \cite{Muzicar}, its appearance in our model  and its dependence on $\kappa$ would benefit from further investigation. One clue could come from the observation made earlier, that it is $\kappa \theta_1$ that  plays the role of an effective CS coupling on the 2+1 dimensional boundary. But $\theta_1$ is a function of $r$, is dynamically determined and implicitly depends  on the value of $\kappa $. It only tends to a constant at large $r $ as can be seen from Figure 4. So in comparing with the analysis of standard 2+1 dimensional Maxwell-CS vortices we have to bear in mind that we have, effectively, an $r$ dependent CS coupling that is only constant for large $r$.

 Another context in which oscillatory behaviour of a magnetic field is observed is when considering non-relativistic vortices in external magnetic and electric fields. Such theories are relevant in the ZHK model of the quantum Hall effect \cite{ZHK } where vortices occur in a statistical gauge field  governed by a pure CS action. The magnetic CS vortices in this context can exhibit oscillatory or `field reversal' behaviour similar to what we have seen in Figure 2  (see section 4 of \cite{Zhang} for a review).  This system is of course different from the one we have considered in this paper, since there is a distinction between the fictitious or 'statistical' CS gauge field and the physical background magnetic field.  It would be worthwhile to investigate how  such non-relativistic CS vortices could arise in a holographic framework.

\section{Acknowledgements}
The work of  G.T. is supported by an EPSRC studentship. We thank David Lin for help in using and providing access to the COMSOL software package.
\section{Appendix}

The $AdS/CFT$ dictionary states that each term in the expansion of the fields at $z\rightarrow0$ is a source for a conformal field at the boundary. More precisely, for a conformal field $<O_i>$ sourced by terms in the expansion of $A_i$

\be
<O_i (x)>=
\frac{1}{\beta\mathcal{V}}\lim_{z\rightarrow0}
\frac{\delta S_{on-shell}}{\delta A_i(x,z)}
\ee

where $x$ are boundary coordinates in the dual theory,  $S_{on-shell}$ is the on shell, Euclidean  version of the  action eq(\ref{action1}), $\beta$ is the inverse temperature and $\mathcal{V}$ is the spatial volume of the dual theory. The calculation to derive the Maxwell contribution to the conformal fields is illustrated in \cite{Johnson}. In our case we have in addition the $\theta F \wedge F$ term which contributes a CS  like term  $2\theta_0\epsilon_{z\mu\rho\sigma}A_{\mu}F_{\rho\sigma} $
when one considers boundary contributions to the variation $\delta S_{on-shell} $.

Hence, including variations of this term with respect to progressive orders in the expansion of $A_i$ at the boundary $z=0$ determine the full conformal fields. This gives,
\bea
\rho_{AdS}&=&-\frac{1}{2}\partial_z A_0 +4\theta_0\partial_rA_\chi\nonumber\\
J^\chi_{AdS}&=&\frac{1}{2}\partial_z\tilde{A}_\chi -4\theta_0\partial_rA_0\nonumber\\
\eea

where importantly we see the non-vanishing contribution of the effective CS terms in the definition of the charge density and azimuthal current. In the above we have ignored the contribution for the black hole horizon at $z=1$ in evaluating
$\delta S_{on-shell} $.


\begin{thebibliography}{1}
\bibitem{Hartnoll}
S.A. Hartnoll, C.P. Herzog, G.T. Horowitz ``Holographic Superconductors" JHEP0812:015,2008
\bibitem{Hartnoll2}
S.A. Hartnoll, C.P. Herzog, G.T. Horowitz ``Building an AdS/CFT superconductor" Phys.Rev.Lett.101:031601,2008
\bibitem{Horowitz}
G.T. Horowitz ``Introduction to Holographic Superconductors", arXiv:1002.1722v2 [hep-th]
\bibitem{Roberts3}
M.M. Roberts, S.A. Hartnoll ``Pseudogap and time reversal breaking in a holographic superconductor" JHEP0808:035,2008
\bibitem{Johnson}
T.Albash, C.V.Johnson,``Vortex and Droplet Engineering in Holographic Superconductors" 10.1103/PhysRevD.80.126009
\bibitem{Johnson2}
T.Albash, C.V.Johnson,``A Holographic Superconductor in an External Magnetic Field" JHEP0809:121,2008

\bibitem{Hartnoll3}
S.A. Hartnoll,``Lectures on holographic methods for condensed matter physics"  arXiv:0903.3246v2 [hep-th]
\bibitem{Pu}
S. Pu, S. Sin, Y. Zhou ``A Holographic model for Non-Relativistic Superconductor" arXiv:0903.4185v3 [hep-th]
\bibitem{Pomarol}
M. Montull, A. Pomarol, P.J.Silva,``The Holographic Superconductor Vortex" arxiv: 0906.2396v3 [hep-th]
\bibitem{Basu}
P.Basu, J. He, A. Mukherjee, H.Shieh``Hard-gapped Holographic Superconductors"  arXiv:0911.4999v2 [hep-th]
\bibitem{Johnson3}
T. Albash, C.V. Johnson,``Phases of Holographic Superconductors in an External Magnetic Fied"	arXiv:0906.0519v1 [hep-th]
\bibitem{Zeng}
H. Zeng, Z. Fan, Z. Ren ``Time Reversal Symmetry Breaking Holographic Superconductor in Constant External Magnetic Field" Phys.Rev.D80:066001,2009
\bibitem{Russo}
F. Aprile, J.G. Russo,``Models of Holographic Superconductivity"  arXiv:0912.0480v1 [hep-th]
\bibitem{Sonner1}
J. Sonner,``A rotating Holographic Superconductor" Phys.Rev.D80:084031,2009
\bibitem{Jerome1}
J.P. Gauntlett, J. Sonner, T. Wiseman,``Holographic Superconductivity in M-Theory" Phys.Rev.Lett.103:151601,2009
\bibitem{Jerome2}
J.P. Gauntlett, J. Sonner, T. Wiseman,``Quantum Criticality and Holographic Superconductors in M-Theory" arXiv:0912.0512v2 [hep-th]
\bibitem{Pan2}
Q. Pan, B. Wang, E. Papantonopoulos, J. Oliveira, A.B. Pavan ``Holographic Superconductors with various condensates in Einstein-Gauss-Bonnet gravity" arXiv:0912.2475v2 [hep-th]
\bibitem{keranen1}
Ville Keranen, Esko Keski-Vakkuri, Sean Nowling, K. P. Yogendran,``Inhomogeneous Structures in Holographic Superfluids: I. Dark Solitons", arXiv:0911.1866v1 [hep-th]
\bibitem{keranen2}
Ville Keranen, Esko Keski-Vakkuri, Sean Nowling, K. P. Yogendran,``Inhomogeneous Structures in Holographic Superfluids: II. Vortices", arXiv:0912.4280v2 [hep-th]
\bibitem{nakamura}
Shin Nakamura, Hirosi Ooguri, Chang-Soon Park. "Gravity Dual of Spatially Modulated Phase", Phys.Rev.D81:044018 (2010).
\bibitem{Ooguri}
Hirosi Ooguri, Chang-Soon Park, "Holographic End-Point of Spatially Modulated Phase Transition", arXiv:1007.3737v1 [hep-th]

\bibitem{Maeda}
K. Maeda, M. Natsuume, T. Okamura ``Vortex lattice for a holographic superconductor" Phys.Rev.D81:026002,2010
\bibitem{Faulkner}
T. Faulkner, G.T. Horowitz, J McGreevy, M.M. Roberts, D. Vegh ``Photoemission "experiments" on holographic superconductors" JHEP 1003:121,2010
\bibitem{Chen2}
S. Chen, L. Wang, C. Ding, J. Jing ``Holographic superconductors in the AdS black hole spacetime with a global monopole" Nucl. Phys. B 836, 222-231(2010)
\bibitem{Jing}
J. Jing, S. Chen ``Holographic superconductors in the Born-Infeld electrodynamics" Physics Letters B 686 (2010) 68-71
\bibitem{Gubser}
S.S. Gubser, F.D. Rocha, A. Yarom ``Fermion correlators in non-abelian holographic superconductors" 	arXiv:1002.4416v1 [hep-th]
\bibitem{Ge}
X. Ge, B. Wang, S. Wu, G. Yang ``Analytical study on holographic superconductors in external magnetic field" arXiv:1002.4901v1 [hep-th]
\bibitem{Setare}
M.R.Setare, D. Momeni ``Holographic superconductors in Horava-Lifshitz Gravity" arXiv:1003.0376v1 [hep-th]
\bibitem{Chen}
J. Chen, Y. Kao, D. Maity, W. Wen, C. Yeh ``Towards A Holographic Model of D-Wave Superconductors" arXiv:1003.2991v5 [hep-th]
\bibitem{Herzog}
C.P. Herzog,``An Analytic Holographic Superconductor" 	arXiv:1003.3278v1 [hep-th]
\bibitem{Siopsis}
G. Siopsis, J. Therrien ``Analytic calculation of properties of holographic superconductors" arXiv:1003.4275v1 [hep-th]
\bibitem{Diego}
F. Aprile, S. Franco, D. Rodriguez-Gomez, J.G.Russo, ``Phenomenological models of Holographic Superconductors and Hall currents" 	arXiv:1003.4487v3 [hep-th]
\bibitem{Brihaye}
Y. Brihaye, B. Hartmann ``Holographic Superconductors in 3+1 dimensions away from the probe limit" 	arXiv:1003.5130v2 [hep-th]
\bibitem{Arean}
D. Arean, M. Bertolini, J. Evslin, T. Prochazka ``On Holographic Superconductors with DC Current" arXiv:1003.5661v2 [hep-th]
\bibitem{Murata}
K. Murata, S. Kinoshita, N. Tanahashi ``Non-equilibrium Condensation Process in a Holographic Superconductor" arXiv:1005.0633v1 [hep-th]
\bibitem{Pomarol2}
Oriol Domènech, Marc Montull, Alex Pomarol, Alberto Salvio, Pedro J. Silva, ``Emergent Gauge Fields in Holographic Superconductors'' arxiv:1005.1776 [hep-th]
\bibitem{Wang}
Q. Pan, B. Wang,``General holographic superconductor models with Gauss-Bonnet corrections" arXiv:1005.4743v1 [hep-th]
\bibitem{Wu}
J. Wu ``The Stuckelberg Holographic Superconductors in Constant External Magnetic Field" 	arXiv:1006.0456v3 [hep-th]
\bibitem{Liu}
Yan Liu, Ya-Wen Sun``Holographic Superconductors from Einstein-Maxwell-Dilaton Gravity" arxiv:1006.2726v2 [hep-th]
\bibitem{arean}
Daniel Arean, Pallab Basu, Chethan Krishnan,``The Many Phases of Holographic Superfluids", arXiv:1006.5165v1 [hep-th].
\bibitem{Zeng}
H. Zeng, Z. Fan, H. Zong ``Characteristic length of a Holographic Superconductor with $d$-wave gap" 	arXiv:1006.5483v1 [hep-th]
\bibitem{Myung}
Y.S. Myung, C. Park ``Holographic superconductor in the exact hairy black hole"  arXiv:1007.0816v1 [hep-th]
\bibitem{horowitz4}
Gary T. Horowitz, Benson Way
'Complete Phase Diagrams for a Holographic Superconductor/Insulator System', arXiv:1007.3714v1 [hep-th]
\bibitem{Gibbons}
G. W. Gibbons, S. W. Hawking, ``Classification Of Gravitational Instanton Symmetries" Commun. Math. Phys. 66 (1979) 291-310.
\bibitem{Nielsen}
H. B. Nielsen , P. Olesen, ``Vortex-line models for dual strings'',  Nucl. Phys, B61: 45, (1973). 
\bibitem{HKP}
J. Hong, Y. Kim, P.Y. Pac,  ``Multivortex solutions of the Abelian Chern-Simons-Higgs theory'', Phys. Rev. Lett. 64, 2230, 1990)
\bibitem{Jackiw}
R. Jackiw, E.J. Weinberg, ``Self-Dual Chern-Simons Vortices" Phys. Rev. Lett. 64, 2234 (1990)
\bibitem{Zhang}
P.A. Horvathy, P. Zhang,``Vortices in (abelian) Chern-Simons gauge theory" Physics Reports 481 83 (2009)
\bibitem{Breiten}
P. Breitenlohner, D.Z. Freedman, ``Positive Energy in anti-De Sitter Backgrounds and Gauged Extended Supergravity", Phys.Lett. B115 (1982) 197.
\bibitem{Reuter}
M. Reuter, ``A mechanism generating axion hair for Kerr black holes'',  Class. Quantum Grav. 9 751 (1992)
\bibitem{Duncan}
Malcolm J. Duncan, N. Kaloper, Keith A. Olive, ``Axion hair and dynamical torsion from anomalies'',  Nucl.Phys.B387:215,(1992).
\bibitem{Maldacena}
J.M. Maldacena ,``The Large N limit of superconformal field theories and supergravity". Advances in Theoretical and Mathematical Physics 2: 231, (1998).
\bibitem{TT}
Gianni Tallarita, Steven Thomas, work in progress.
\bibitem{Muzicar}
P. Muzicar, ``Vortex lines and field reversal in anisotropic superconductors'', Journal of Low Temperature Physics, Volume 115, 3-4 (1999)
\bibitem{ZHK}
S. C. Zhang ,T. H. Hansson, S. Kivelson , ``Effective-Field-Theory Model for the Fractional Quantum Hall Effect'', Phys. Rev. Lett. 62, 82-85 (1989).

\end{thebibliography}
\end{document}